\documentclass[universe,review,accept,oneauthor]{Definitions/mdpi}
\usepackage{amsmath}
\usepackage{amsfonts}
\usepackage{amssymb,bm}
\usepackage{siunitx}
\usepackage{color}

\firstpage{1}
\pubvolume{7}
\issuenum{3}
\articlenumber{47}
\pubyear{2021}
\copyrightyear{2021}
\history{Received: 4 February 2021; Accepted: 23 February 2021; Published: 26 February 2021}

\Title{Constraints on Theoretical Predictions Beyond the Standard Model
from the Casimir Effect and Some Other Tabletop Physics}

\Author{Galina L. Klimchitskaya $^{1,2}$
}

\address{%
$^{1}$ \quad Central Astronomical Observatory at Pulkovo of the Russian Academy of Sciences, Saint Petersburg, 196140, Russia\\
$^{2}$ \quad Institute of Physics, Nanotechnology and
Telecommunications, Peter the Great Saint Petersburg
Polytechnic University, Saint Petersburg, 195251,  Russia}

\corres{Correspondence: g.klimchitskaya@gmail.com}

\abstract{We review the hypothetical interactions predicted
beyond the Standard Model which could be constrained by using
the results of tabletop laboratory experiments. These interactions
are described by the power-type potentials with different powers,
Yukawa potential, other spin-independent potentials, and by the
spin-dependent potentials of different kinds. In all these cases
the current constraints on respective hypothetical interactions
are considered which follow from the Casimir effect and some
other tabletop physics. The exotic particles and constraints on
them are discussed in the context of problems of the quantum
vacuum, dark energy, and the cosmological constant.
}


\begin{document}

\section{Introduction}

Despite great successes of the Standard Model, there is general
agreement that it does not provide a solution to a number of
crucial problems of modern physics such as quantization of the
gravitational field \cite{1}, the enormously large energy density
of the quantum vacuum \cite{2}, strong CP violation in QCD \cite{3},
the problems of dark matter and dark energy \cite{4}, etc. In
attempting to solve these fundamental problems in the framework of
the extended Standard Model, supersymmetry, supergravity and
string theory, a lot of hypothetical interactions and elementary
particles have been introduced which are characterized by a very
weak interaction with particles of the Standard Model and cannot
be detected using the present day accelerator techniques.

Among these particles are the massless pseudoscalar Goldstone
bosons arions \cite{5}, light pseudo Nambu-Goldstone bosons
axions \cite{6,7} which arise from violation of the Peccei-Quinn
symmetry \cite{8}, spin-1 elementary excitations of the
gravitational field in extra dimensions graviphotons \cite{9},
scalar particles dilatons (radions) which arise in the
multidimensional schemes with spontaneously compactified extra
dimensions \cite{10}, the Nambu-Goldstone fermions
goldstinos introduced in the supersymmetric theories with a
spontaneously broken supersymmetry \cite{11,12}, etc. During
the last few years much attention was attracted to the
self-interacting scalar particles chameleons having a variable
mass which is smaller in the regions of small matter density
(i.e., in the interstellar space) and larger in the regions of
large matter density (e.g., on the Earth) \cite{13}. In a similar
manner, the scalar fields called symmetrons have arosed
considerable interest whose coupling constant depends on the
density of matter in the environment \cite{14}.

An exchange of light and massless particles between the microscopic
constituents of macroscopic bodies results in initiation of the
effective interaction potentials and forces which are additional
to the previously known fundamental forces, i.e, the
gravitational, electromagnetic, weak and strong interactions.
Any new force of this kind independently of its origin is often
called {\it the fifth force} \cite{15}. There are different types
of effective potentials describing the new forces. For instance,
an exchange of massless hypothetical particles of different
spins leads to the power-type potentials and forces which are
inverse proportional to different powers of separation. Note that
the power-type corrections to Newton's gravitational law also
arise in multidimensional theories with noncompact but warped
extra dimensions \cite{16,17}. An exchange of light scalar
particles leads to the most popular new forces of Yukawa type
\cite{15}, whereas an exchange of one preudoscalar particle
between nucleons of two macrobodies leads to the
spin-dependent effective potential \cite{18,19,20}. The
Yukawa-type corrections to Newtonian gravity accessible to
observations are also predicted by the multidimensional theories
with compact extra dimensions and relatively low compactification
scale of the order of 1 TeV \cite{21,22,23,24}. The more
complicated spin-independent interactions originating from an
exchange of two axions and other hypothetical particles have
also been predicted (see, e.g., \cite{19,20,25}.

As already noted, it is unlikely that the predicted new forces
will be observed using the accelerator techniques. They lead,
however, to some additional interactions between the closely
spaced macroscopic bodies which contain a huge number of
elementary constituents and, thus, could be searched for in the
tabletop laboratory experiments. These are the gravitational
experiments of E\"{o}tvos and Cavendish type, experiments on
neutron scattering, magnetometer measurements, measuring the
decay rates of hypothetical particles into photons in strong
magnetic field, precision atomic physics, etc. (see the reviews
\cite{26,27,28,29,30}.

Starting from the pioneer paper \cite{31}, experiments on
measuring the van der Waals and Casimir forces are succesfully
used for constraining the hypothetical forces of Yukawa type.
The point is that at separations below a few micrometers
between the test bodies the gravitational interaction becomes
so small that even by the orders of magnitude larger additions
to it are not experimentally excluded. Thus, when searching
for some hypothetical force at so short separations, one
needs to use another familiar background force whose deviations
from its theoretical law could serve as an alarm that one more
physical effect comes into play. The van der Waals and Casimir
forces, which are caused by the zero-point and thermal
fluctuations of the electromagnetic field, just form the
background of this kind in the separation range from a nanometer
to a few micrometers. Therefore, any precise measurement of
these forces can be used to impose constraints on some
hypothetical interaction relevant to the same range of
separations.

This article reviews the laboratory constraints on various
hypothetical particles and interactions beyond the Standard
Model obtained from experiments on measuring the Casimir
interaction and some other tabletop experiments which lead to
constraints in the parameter regions neighboring to those
covered by the Casimir effect. We start from constraints on the
power-type interactions with different powers and continue with
the interactions of Yukawa type which are often considered in
the literature as corrections to Newton's law of gravitation
at short separations. Next we deal with some other
spin-independent effective potentials which describe an
exchange of two axionlike particles and certain other
processes. Special attention is given to the constraints on
the spin-dependent interactions originating from an exchange
of one axion and by several more exotic processes. Axions
and axionlike particles are often considered as possible
constituents of dark matter \cite{27,29}. Constraints on the parameters
of several exotic particles like chameleons, symmetrons and
massive photons are also considered, as well as their
implications to the problem of dark energy. In some cases
not only the already obtained constraints are presented but
various proposals, plans and suggestions allowing their
strengthening are discussed as well.
Note that many corrections to the Casimir force which are a priori much
smaller than the corrections due to nonideality of the plate metal are
widely discussed in the literature (see, e.g., the corrections in theories
with a minimal length \cite{31a} or in a static space-time with a
Lorenz-violating term \cite{31b}). All these subjects are outside the scope
of our review.

The article is organized as follows: In Sections 2 and 3 the
interactions of the power and Yukawa type are considered,
respectively. Section 4 is devoted to other spin-independent
interactions. In Section 5, the constraints on spin-dependent
interactions are discussed. Sections 6 and 7 deal with some
exotic particles and their implications to the problems of
quantum vacuum, dark energy and the cosmological constant. In
Section 8 the reader will find the discussion and in Section
9 -- our conclusions. We use the system of units where
$\hbar=c=1$.

\section{The Hypothetical Interactions of Power Type}

Potentials of the hypothetical interactions of power
and Yukawa type are usually presented in the form of
corrections to the Newtonian gravitational potential.
Thus, the power-type corrections can be parametrized as
\begin{equation}
V_l(r)=V_N(r)\left[1+\Lambda_l\left(\frac{r_0}{r}\right)^{l-1}\right],
\qquad
V_N(r)=-\frac{Gm_1m_2}{r},
\label{eq1}
\end{equation}
\noindent
where $G$ is the gravitational constant, $m_1$ and $m_2$ are the two point masses
(atoms, nucleons) spaced at the points {\boldmath$r_1$} and {\boldmath$r_2$},
$r=|\mbox{\boldmath$r_1$}-\mbox{\boldmath$r_2$}|$, and $\Lambda_l$ with
$l=1,\,2,\,3,\,\ldots$ are the dimensionless constants characterizing the strength
of the power-type interaction. The quantity $r_0$ is introduced to preserve the
correct dimension of $V_l$. It is often chosen as $r_0=1~\mbox{F}=10^{-15}~$m.
Note that (\ref{eq1}) is unrelated to the post-Newtonian approach presenting the
deviations of Einstein's general relativity theory from Newton's law in powers
of some small parameter. Here, the correction to unity can be of nongravitational
origin.

The effective potential (\ref{eq1}) with $l=1$ arises due to an exchange of one
massless particle (the Coulomb-type potential). An exchange of two arions between
electrons belonging to atoms of two neighboring macrobodies results in the
power-type correction with $l=3$ which decreases with separation as $r^{-3}$
\cite{32}. The power-type corrections to the Newtonian potential with higher
powers result from an exchange of the even numbers of neutrinos, goldstinos, and
other massless fermions. For instance, an exchange of the
neutrino-antineutrino pair between two neutrons leads to the power-type
correction with $l=5$  which decreases with increasing separation as $r^{-5}$
\cite{32a,33}. For the multidimensional models with noncompact extra dimensions
at separations $r\gg 1/k$, where $k$ is the warping scale, the effective
potential takes the form of (\ref{eq1}) with $l=3$,
$\Lambda_3=2/(3k^2r_0^2)$ \cite{16,17}.

The constraints on the interaction constants $\Lambda_l$ with different $l$ can
be obtained from the E\"{o}tvos-type or Cavendish-type experiments.
The presence of some hypothetical interactions in addition to the Newtonian
gravitation would lead to a seeming difference between the inertial and gravitational
masses, i.e., to a violation of the equivalence principle tested by the
E\"{o}tvos-type  experiments. Using this approach, the maximum values of
$|\Lambda_1|$ and  $|\Lambda_2|$ allowed by the short-range E\"{o}tvos-type
experiments are the following:
\begin{equation}
|\Lambda_1|_{\max}=1\times 10^{-9}{\ \ }\mbox{\cite{34}},
\qquad
|\Lambda_2|_{\max}=4\times 10^8{\ \ }\mbox{\cite{35}}.
\label{eq2}
\end{equation}
\noindent
We recall that even a confirmed difference of $\Lambda_l$ from zero would not imply
a violation of the equivalence principle if the correction term in (\ref{eq1}) is
of nongravitational origin.

As to the Cavendish-type experiments, they test deviations of the gravitational
potential $V_{\rm gr}(r)$ from Newton's law. These deviations can be quantified
by the value of the dimensionless parameter
\begin{equation}
\delta=\frac{1}{V_{\rm gr}(r)}\frac{\partial}{\partial r}\left[
rV_{\rm gr}(r)\right],
\label{eq3}
\end{equation}
\noindent
which is identically equal to zero if $V_{\rm gr}(r)=V_{N}(r)$ defined in (\ref{eq1}).
On this basis, the maximum values of $|\Lambda_l|$ with $l=2,\,3,\,4$ and 5 allowed
by the Cavendish-type experiment performed in 2007 were found to be \cite{36}
\begin{equation}
|\Lambda_2|_{\max}=4.5\times 10^{8},
\quad
|\Lambda_3|_{\max}=1.3\times 10^{20},
\quad
|\Lambda_4|_{\max}=4.9\times 10^{31},
\quad
|\Lambda_5|_{\max}=1.5\times 10^{43}.
\label{eq4}
\end{equation}
\noindent
One should take into account that we have recalculated the results of \cite{36}
to our choice of $r_0$ (in \cite{36} $r_0=1~$mm was used).
As is seen  from (\ref{eq2}), the constraint on $|\Lambda_2|$ in (\ref{eq4}) is
slightly weaker than that one following from the E\"{o}tvos-type  experiment
\cite{35}.

In \cite{32,37} in was suggested to constrain $\Lambda_l$ using the measure
of agreement between experiment and theory for the Casimir force measured in
\cite{38} in the configuration of a plate and a spherical lens.
It was shown \cite{32,37} that measurements of the Casimir force result in
\begin{equation}
|\Lambda_2|_{\max}=1.7\times 10^{11},
\quad
|\Lambda_3|_{\max}=8.5\times 10^{23}.
\label{eq5}
\end{equation}
\noindent
These constraints are weaker than the constraints (\ref{eq2}) and (\ref{eq4})
following from the experiments of E\"{o}tvos and Cavendish type.
In 1987, however, when the constraints (\ref{eq5}) on $|\Lambda_2|$ and
$|\Lambda_3|$ from the Casimir effect have been obtained, they were the
strongest ones. In succeeding years there was some kind of competition
between measurements of the Casimir force and gravitational experiments in
obtaining the strongest constraints on $|\Lambda_l|$ with $l=2$ and 3
\cite{39,40}.

The constraints on $|\Lambda_1|$ and $|\Lambda_2|$ in (\ref{eq2}) and on
$|\Lambda_l|$ with $l=2,\,3,\,4$ and 5 in (\ref{eq4}) following from the
E\"{o}tvos- and Cavendish-type experiments were the strongest ones during
the period from 2007 to 2020. In 2020 the new Cavendish-type experiment
has been performed \cite{41} providing an improved test of the Newton law
at short separations. As a result, somewhat stronger constraints on
$|\Lambda_l|$ with $l=2,\,3,\,4$, and 5 than in (\ref{eq4}) were obtained.
In Table~1 we summarize the strongest current constraints on $\Lambda_l$.
\begin{table}[H]
\caption{The strongest current constraints on the potentials of power type.}
\centering
\begin{tabular}{ccc}
\toprule
\textbf{$l$}	& \textbf{$|\Lambda_l|_{\max}$}	& \textbf{Source}\\
\midrule
1		& $1\times 10^{-9}$			& \cite{34}\\
2		& $3.7\times 10^{8}$	    & \cite{41}\\
3		& $7.5\times 10^{19}$	    & \cite{41}\\
4		& $2.2\times 10^{31}$	    & \cite{41}\\
5		& $6.7\times 10^{42}$	    & \cite{41}\\
\bottomrule
\end{tabular}
\end{table}

The probability exists of further strengthening of the constraints of Table~1. By way
of example, there are proposals in the literature to constrain the
power-type hypothetical interactions of different origin at micrometer
separations by means of optomechanical methods exploiting the levitated sensors
\cite{42} or using experiments on neutron scattering and molecular spectroscopy
\cite{43}.

\section{The Yukawa-Type Interactions}

The potential of Yukawa type is commonly presented as a correction to Newtonian
gravitation
\begin{equation}
V(r)=V_N(r)\left(1+\alpha\,e^{-\frac{r}{\lambda}}\right),
\label{eq6}
\end{equation}
\noindent
where $\alpha$ is the dimensionless constant characterizing the interaction strength
and $\lambda$ is the interaction range. As mentioned in Section~1, the Yukawa-type
correction to Newton's law arises due to an exchange of one light scalar particle
of mass $m=1/\lambda$ between atoms of two macroscopic bodies and in
multidimensional theories with compact extra dimensions which are compactified at
the low energy scale. This motivated numerous experiments searching for the corrections
of  this kind.

When integrated over the volumes $V_1$ and $V_2$ of two test bodies spaced at a small
distance $a$, the potential (\ref{eq6}) results in the interaction energy of these
bodies. Calculating the negative derivative of this energy with respect to $a$,
one arrives to the force
\begin{equation}
F(a)=-\frac{1}{V_1V_2}\,\frac{\partial}{\partial a}\int_{V_1}\int_{V_2}\!
V(r)d^{\,3}r_1d^{\,3}r_2.
\label{eq7}
\end{equation}
\noindent
At separations $a$ below a few micrometers the contribution of the Newtonian gravity
to (\ref{eq7}) is usually negligibly small as compared to the sensitivity of force
measurements.

As was noted in Section~1, at separations below a few micrometers the main background
force acting between electrically neutral test bodies is the Casimir force caused
by the zero-point and thermal fluctuations of the electromagnetic field \cite{44}.
In the beginning of the XXI century the Casimir force and its gradient were measured
in many precision experiments (see \cite{45} for a review). The measurement
results were found in agreement with theoretical predictions of the fundamental
Lifshitz theory in the limits of some errors. By assuming that the Casimir forces
$F_C^{\rm expt}(a_i)$ measured at the separations $a_i$ agree with the theoretical
predictions $F_C^{\rm theor}(a_i)$ up to the errors $\Delta_iF_C$, i.e., that
the inequality
\begin{equation}
\left|F_C^{\rm expt}(a_i)-F_C^{\rm theor}(a_i)\right|\leqslant\Delta_iF_C
\label{eq8}
\end{equation}
\noindent
is satisfied, one arrives at a conclusion that any additional force (\ref{eq7})
must satisfy the condition
\begin{equation}
|F(a_i)|\leqslant\Delta_iF_C.
\label{eq9}
\end{equation}
\noindent
This condition imposes some constraints on the parameters  $\alpha$ and
$\lambda$ of the Yukawa interaction defined in (\ref{eq6}) and (\ref{eq7}).

This approach, as mentioned in Section~1, was first used in \cite{31}
where the constraints on the Yukawa-type interaction were obtained from old
measurements of the van der Waals force \cite{38,46}. According to the results
of \cite{31}, measurements of the van der Waals interaction place the
strongest constraints on scalar particles with $m>10^{-6}~$eV.
This corresponds to the interaction range $\lambda<20~$cm.
At larger $\lambda$ (smaller masses) the strongest constraints on the
Yukawa interaction follow from gravitational experiments \cite{31}.
Below it is shown that modern experiments on measuring the Casimir force,
as well as the new tests of the inverse-square law of gravity and
equivalence principle, significantly modify these results.

During the last 20 years, many measurements of the Casimir force have been used
to obtain constraints on the Yukawa-type interaction. Thus, in \cite{47}  the
competitive constraints at small $\lambda$ were found from measuring the
Casimir force between two crossed cylinders \cite{48}. These constraints have
been further confirmed and strengthened in \cite{49,50,51} using experiments
on measuring the lateral and normal Casimir forces between sinusoidally
corrugated surfaces \cite{52,53,54,55} and the normal Casimir force in the
configuration of a silicon carbide surface situated at only 10~nm
separation from a sphere \cite{56,57}. At larger $\lambda$ strong constraints
on the strength of Yukawa interaction $\alpha$ were obtained from measuring
the effective Casimir pressure by means of micromechanical torsional
oscillator \cite{58,59,60,61}. These results were confirmed \cite{62}
using the experimental data of experiment on measuring the difference of
Casimir forces \cite{63}. Major progress in the field was reached by
the so-called {\it Casimir-less} experiments which also measure the
differential forces and are organized in such a way that the contribution
of the Casimir force is completely nullified \cite{64,65}. These
experiments are especially sensitive to the presence of any additional
interaction. The second of them \cite{65} resulted in the most strong
constraints over a wide interaction range. Much weaker constraints on
$\alpha$ at relatively large $\lambda$ were obtained in experiments using
the torsion pendulum \cite{66} and measuring the difference in the
lateral forces \cite{67}.

\begin{figure}[!t]
\centering
\vspace*{-8.cm}
\includegraphics[width=18 cm]{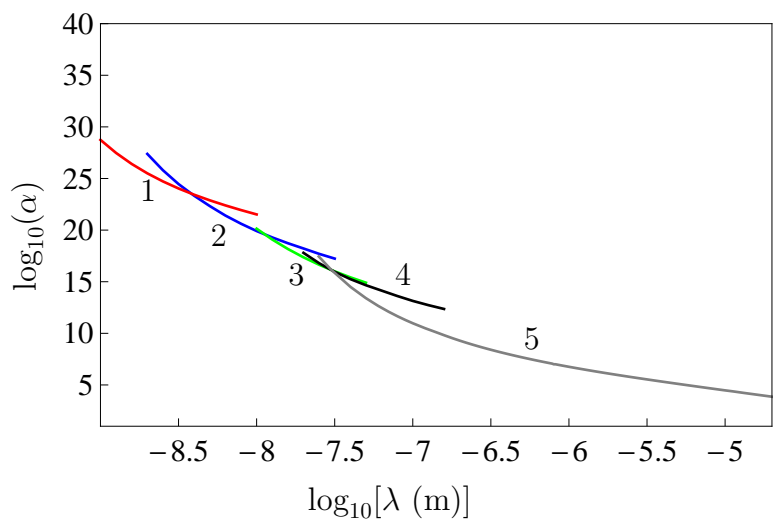}
\vspace*{-10.5cm}
\caption{The strongest constraints on the Yukawa-type potentials obtained
from measuring the Casimir force at nanometer separations (line 1),
 lateral  Casimir force between corrugated surfaces (line 2),
 normal Casimir force between corrugated surfaces (line 3),
 the Casimir pressure (line 4), and from the  Casimir-less experiment
(line 5).
The plane areas above the lines are excluded and below them
are allowed by the measurement data.\label{fg1}}
\end{figure}
In more detail, the above-listed experiments and constraints obtained from
them are reviewed in \cite{68}.
In Figure~\ref{fg1}, we present in the logarithmic scale only the strongest constraints
on $\alpha$ for different $\lambda>1~$nm obtained from measurements of the
Casimir force. The line 1 is found \cite{51} from \cite{56,57},
line 2  \cite{49} from \cite{52,53}, line 3  \cite{50} from \cite{54,55},
line 4  is obtained in  \cite{60,61}, and line 5 in \cite{65}.
The $(\lambda,\,\alpha)$-plane areas above each line are excluded by the
measurement data of respective experiment and the plane areas below each line
are allowed. To summarize, the line 1 presents the strongest constraints
in the interaction region $1~\mbox{nm}\leqslant\lambda<3.7~$nm,
the line 2 in the region $3.7~\mbox{nm}\leqslant\lambda<11.6~$nm,
the line 3 in the region $11.6~\mbox{nm}\leqslant\lambda<17.2~$nm,
the line 4 in the region $17.2~\mbox{nm}\leqslant\lambda<40~$nm,
and the line 5 for $\lambda\geqslant 40~$nm.

Similar to the power-type potentials, the potentials of Yukawa type are
constrained by the results of Cavendish- and E\"{o}tvos-type experiments.
The gravitational experiments are the most sensitive to the presence of
additional interactions if the separation distance between the test bodies is
sufficiently large so that gravitation remains to be the main background
force. The most precise short-range Cavendish-type experiment of this kind
was reported in \cite{69}. For the interaction range of $\lambda$
exceeding $8~\mu$m the constraints on $\alpha$ following from \cite{69}
are stronger than those obtained in \cite{65} from the Casimir physics.
Thus, one can say that  $\lambda=8~\mu$m (the respective mass of the
hypothetical scalar particle is $m=1/\lambda\approx 2.5\times 10^{-2}~$eV)
is the upper border of the current region of $\lambda$ where the Casimir
physics provides the strongest constraints on the Yukawa-type potentials.
This means that during the period of time after 1987, when the possibility
to constrain the Yukawa-type potentials from the Casimir effect was
proposed \cite{31}, the role of gravitational and Casimir experiments in
obtaining stronger constraints has vastly changed.

The constraints of \cite{69} are the strongest ones up to $\lambda=9~\mu$m.
Within the wide interaction range from $\lambda=9~\mu$m to $\lambda=4~$mm
the strongest constraints on the Yukawa-type potentials follow \cite{36} from
another Cavendish-type experiment \cite{70}. For even larger $\lambda$ up to
1~cm the stronger constraints on $\alpha$ follow from an older Cavendish-type
experiment \cite{71} wherein the test masses were located at larger
separations. It should be mentioned also that in 2020 the constraints on
$\alpha$ found in \cite{70} were strengthened by up to a factor of 3 over
the interaction range $40~\mu\mbox{m}\leqslant\lambda<350~\mu$m by
refining the present techniques of Cavendish-type experiments \cite{41}
(as mentioned in Section 2, this experiment also leads to stronger
constraints on the power-type potentials). For $\lambda>1~$cm the best
test for non-Newtonian gravity of Yukawa type is provided by the
E\"{o}tvos-type experiments \cite{35,72}.

\begin{figure}[!t]
\centering
\vspace*{-6.5cm}
\includegraphics[width=18 cm]{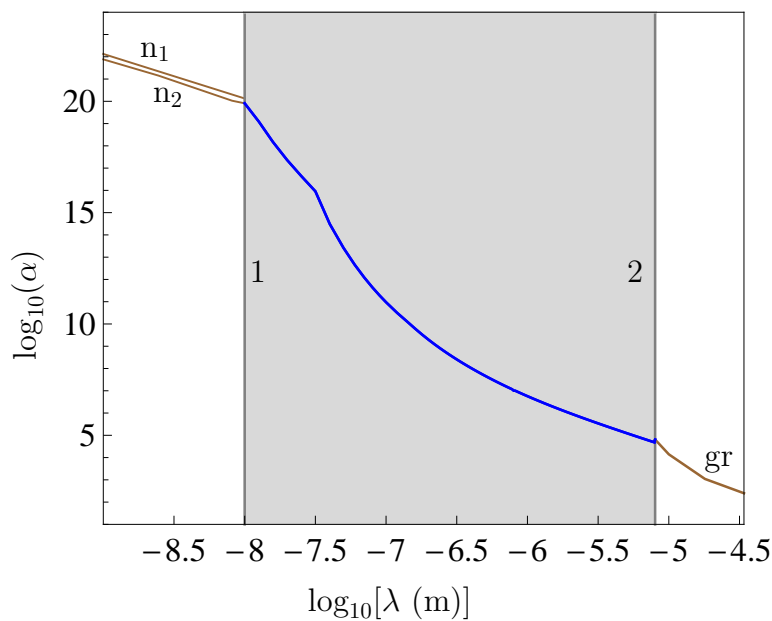}
\vspace*{-10.5cm}
\caption{The strongest constraints on the Yukawa-type potentials obtained
from two Cavendish-type experiments (line labeled "gr") and two
experiments on neutron scattering (lines labeled "n$_1$" and
"n$_2$"). The vertical lines labeled 1 and 2 indicate the current
borders of the constraints obtained from Casimir physics.
The latter are shown by the envelope curve in between the vertical
lines 1 and 2.
The plane areas above the lines are excluded and below them
are allowed by the measurement data.\label{fg2}}
\end{figure}
We illustrate the relative role of Casimir and gravitational experiments in
strengthening the Yukawa-type interactions in Figure~\ref{fg2}.
The line marked gr in Figure~\ref{fg2} shows the constraints on the
Yukawa-type potential from the Cavendish-type experiments \cite{69,70}.
In doing so, the constraints of the experiment \cite{70} are shown from
$\lambda=9~\mu$m to $\lambda=34~\mu$m. The vertical line 2 at $\lambda=8~\mu$m
indicates the current border between the strongest constraints on the
Yukawa-type potentials obtained from gravitational experiments
($\lambda>8~\mu$m) and from the Casimir effect ($\lambda\leqslant8~\mu$m).
At $\lambda\leqslant8~\mu$m in Figure~\ref{fg2} we plot the envelope curve
of the strongest constraints on the  Yukawa-type potentials from different
Casimir experiments shown in Figure~\ref{fg1}.

In the region of small $\lambda$ the possibilities of Casimir physics in obtaining
the strongest constraints on the Yukawa-type potentials are restricted by
spectroscopic measurements in simple atomic systems like hydrogen or deuterium
and scattering of slow neutrons on atoms. Thus, from the comparison of
spectroscopic measurements with precise QED calculations it was found \cite{72a}
that in the region $2\times 10^{-4}~\mbox{nm}<\lambda<20~$nm the interaction
constant $\alpha$ varies from $2\times 10^{27}$ to $2\times 10^{25}$.
The strongest constraints in the interaction range around 1~nm were, however,
obtained from two neutron experiments \cite{73,74}. In Figure~\ref{fg2}
these constraints are shown by the lines n$_1$ \cite{73} and n$_2$ \cite{74}.
These constraints remain the strongest ones up to $\lambda=10~$nm where they
are replaced with the constraints from measurements of the lateral Casimir
force between corrugated surfaces. Thus, the vertical line 1 in Figure~\ref{fg2}
at $\lambda=10^{-8}~$m separates the constraints on the Yukawa-type potentials
obtained from the neutron scattering and from the Casimir physics.
Note that the constraints of \cite{73,74} remain the strongest ones at
$\lambda<1~$nm as well (in the interaction region $0.03~\mbox{nm}<\lambda<0.1~$nm
they have been further strengthened by the experiment using a pulsed neutron
beam \cite{75}).

One can conclude that at the moment Casimir physics provides the strongest
constraints on the Yukawa-type potentials over the wide interaction range
from $\lambda=10^{-8}~$m to $0.8\times10^{-5}~$m.

In the end of this section, we briefly discuss some suggestions on how the
obtained constraints on the Yukawa-type potentials could be strengthened.
Some improvements of the already performed experiments, which promise an
obtaining up to one order of magnitude stronger constraints, were suggested
in \cite{76}. Along with experiments employing the sphere-plate geometry, the
configurations of sinusoidally corrugated test bodies were considered.
In this case the improvements in sensitivity can be reached by increasing
the corrugation amplitudes and decreasing the corrugation period \cite{76}.
The levitated nanoparticle sensor, which is sensitive to static forces
down to $10^{-17}~$N and could be helpful in constraining the Yukawa-type
interactions, was proposed in \cite{77}. In \cite{78} it was argued that
precision spectroscopy of weakly bound molecules can be used for sensing
the non-Newtonian interactions between atoms. Calculations show that in the
interaction range from $\lambda=2~$nm to 10~nm this method allows to
strengthen the constraints obtained from neutron scattering experiments
by at least 1.5--2 orders of magnitude. The possibility to improve the
constraints on $\alpha$ in the wide interaction region around
$\lambda=1~\mu$m by measuring the Casimir-Polder force in the configuration
of a Rb atom and a movable Si plate with an Au film in between was proposed
in \cite{79}. Up to an order of magnitude stronger constraints on $\alpha$
in the interaction range from below a micrometer to $20~\mu$m is promised
by the proposed CANNEX test (Casimir And Non-Newtonian force Experiment)
which is adapted for measuring the Casimir pressure between two parallel
plates at separations up to $10~\mu$m \cite{80,81}.

It should be noted that although the Yukawa-type corrections to the Newton
gravitational law are short-ranged, they are important not only in the
tabletop laboratory experiments but in astrophysics as well. For instance,
as shown in \cite{82}, these corrections make an impact on the properties of
quark stars and should be taken into account for interpretation of
observations related to some specific events and objects.

\section{Other Spin-Independent Hypothetical Interactions}

The most popular spin-independent effective potential other than that of
the Yukawa type originates from an exchange of two light pseudoscalar
particles between two fermions under an assumption of the pseudoscalar
coupling. This means that the Lagrangian density takes the form
\cite{83,84,85}
\begin{equation}
{\cal L}_{P}(x)=-{\rm i}g\bar{\psi}(x)\gamma_5\psi(x)\varphi(x),
\label{eq10}
\end{equation}
\noindent
where $\psi(x)$ and $\varphi(x)$ are the spinor and pseudoscalar fields,
respectively, $\gamma_5$ is the Dirac matrix and $g$ is the dimensionless
interaction constant.

Under a condition that $r\gg m^{-1}$, where $r$ is the separation distance between
two fermions and $m$ is their mass, the effective potential of their interaction
due to an exchange of two pseudoscalars takes the form \cite{15,19,86}
\begin{equation}
V(r)=-\frac{g^4}{32\pi^3m^2}\,\frac{m_a}{r^2}K_1(2m_ar),
\label{eq11}
\end{equation}
\noindent
where $m_a$ is the mass of a pseudoscalar particle and $K_1(z)$ is the modified
Bessel function of the second kind.

The most important hypothetical pseudoscalar particles are axions which were
introduced \cite{6,7} in order to solve the problems of strong CP violation
and large electric dipole moment of neutron which arise in quantum chromodynamics.
The originally introduced axions are the Nambu-Goldstone bosons
and their interaction with fermions is not described by (\ref{eq10})
(see the next section).
Later, however, some other types of axions (axionlike particles) were introduced,
e.g., the so-called Grand-Unified-Theory (GUT) axions, which are coupled to
fermions according to (\ref{eq10}) \cite{84}. Axions and axionlike particles are
presently considered as the most probable constituents of dark matter and
their search in underway in many laboratories all over the world (see
\cite{27,29,84,85} for a review).

In connection with this, much attention was recently attracted to the effective
potential (\ref{eq11}) which describes some additional force arising between two
test bodies due to an exchange by the pairs of axions between their constituents.
An interaction of two macrobodies by means of the two-axion exchange between
their atomic electrons turned out to be too weak and does not lead to the competitive
constraints on the axion-electron interaction. As to the interaction of axions
with nucleons belonging to atomic nuclei, it leads to the competitive laboratory
constraints on $g$ in the wide range of axion masses $m_a$ obtained from
experiments on measuring the Casimir force and gravitational experiments
(the stronger constraints obtained from astrophysics are of lesser reliability
because the theory of dense nuclear matter is still not clearly understood
\cite{87}).

By integrating (\ref{eq11}) over the volumes of two test bodies $V_1$ and $V_2$
and calculating the negative derivative of the obtained result, one obtains the
additional force which arises due to the two-axion exchange between nucleons
\begin{equation}
F_{an}(a)=-\frac{m_ag^4}{32\pi^3m^2}n_1n_2\frac{\partial}{\partial a}
\int_{V_1}\int_{V_2}d{\mbox{\boldmath$r$}}_1 d{\mbox{\boldmath$r$}}_2
\frac{K_1(2m_ar)}{r^2},
\label{eq12}
\end{equation}
\noindent
where $r=|\mbox{\boldmath$r$}_1-\mbox{\boldmath$r$}_2|$ is a distance between
nucleons belonging to $V_1$ and $V_2$, and $n_{1,\,2}$ are the numbers of nucleons
per unit volume of  $V_1$ and $V_2$ (below we assume equal the coupling constants
of an axion to a proton and a neutron). Then, similar to the interaction of
Yukawa type, one can obtain constraints on $m_a$ and $g$ from the inequality
\begin{equation}
|F_{an}(a)|\leqslant \Delta_iF_C,
\label{eq13}
\end{equation}
\noindent
where $\Delta_iF_C$ is the measure of agreement between the experimental Casimir
forces and theoretical predictions [compare with (\ref{eq8}) and (\ref{eq9})].

The first constraints of this kind on the effective potential (\ref{eq11}) and
respective parameters of axionlike particles $m_a$ and $g$ were obtained \cite{88}
from measuring the Casimir-Polder force between the ${}^{87}$Rb atoms and a silica
glass plate \cite{89}. Somewhat stronger constraints in the region of axion masses
below 1~eV were found \cite{90} from the experiment on measuring the gradient of
the Casimir force between Au-coated surfaces of a sphere and a plate performed
by means of an atomic force microscope \cite{91}. All other Casimir experiments
used for constraining the interaction of axionlike particles with nucleons are
the same as already discussed in Section~3 in connection with constraining the
Yukawa-type interaction. Thus, measurements of the effective Casimir pressure
by means of micromechanical torsional oscillator \cite{60,61} were applied for
obtaining stronger constraints on axionlike particles in the region of axion
masses $m_a\geqslant 1~$eV \cite{92}. Measurements of the lateral Casimir force
between sinusoidally corrugated surfaces \cite{52,53} resulted in stronger
constraints \cite{93} for somewhat larger $m_a$ than the constraints of \cite{92}
obtained from \cite{60,61}. The most strong constraints in the range of axion
masses $m_a<1~$eV were obtained \cite{94} from the second Casimir-less experiment
\cite{65}. These constraints were confirmed by slightly weaker independent
constraints derived \cite{62} from measuring the difference of Casimir forces
\cite{63}. Very recently, strong constraints on $g$ in the region of axion
masses up to $m_a=100~$eV were obtained \cite{51} from the experiment where the
Casimir force was measured between a silicon carbide plate at 10~nm minimum
separation from a metallic sphere \cite{56,57}.

Similar to the Yukawa-type correction to Newtonian gravity, the additional
interaction caused by the effective potential (\ref{eq11}) describing two-axion
exchange leads to a difference between the inertial and gravitational masses
and to deviations of the total force between two macrobodies from the inverse
square law. This means that the presence of the force (\ref{eq12}) can be tested in
gravitational experiments. Following this line of attack, the constraints on $g$
in the region of axion masses from $m_a\sim 10^{-8}~$eV to  $m_a\sim 10^{-5}~$eV
were obtained in \cite{20} from the E\"{o}tvos-type experiment \cite{35}.
Somewhat stronger constraints in a wider range of axion masses were found \cite{20}
from the Cavendish-type experiments \cite{71,95}. However, the strongest laboratory
constraints on the coupling constant of axionline particles to nucleons over the
wide range of axion masses were obtained in \cite{36} from the measurement results of
more recent Cavendish-type experiment \cite{70} which was already used in Section~3
for constraining the Yukawa-type interactions.

For slightly larger axion masses stronger constraints on $g$ were obtained \cite{25}
from the experiment on measuring the minimum forces of gravitational strength using
the planar torsional oscillators \cite{96,97,98}.

It should be recalled that the constraints on $m_a$ and $g$ following from the Casimir
physics and from the experiments of E\"{o}tvos and Cavendish type are only appropriate
for the GUT axions whose interactions with nucleons are described by the Lagrangian
density (\ref{eq10}). All these constraints are obtained using the effective potential
(\ref{eq11}) describing the two-axion exchange. Because of this, it is reasonable to
consider the constraints of this type together.

\begin{figure}[!t]
\centering
\vspace*{-6.5cm}
\includegraphics[width=18 cm]{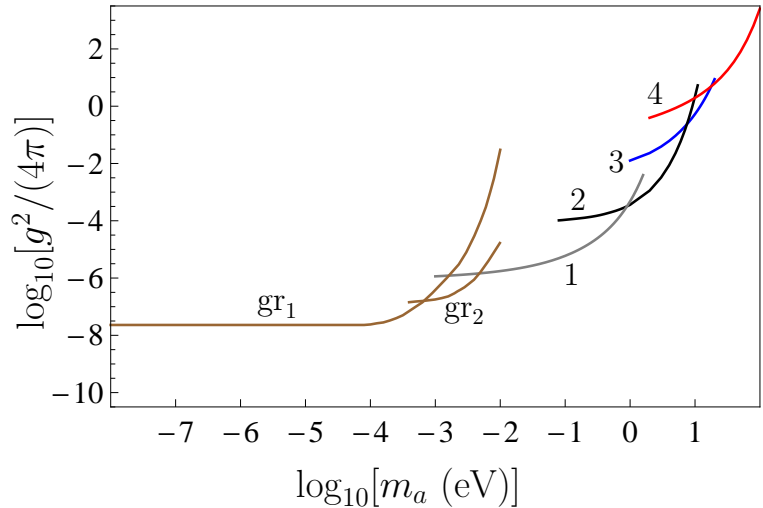}
\vspace*{-10.5cm}
\caption{The strongest laboratory constraints on the
coupling constant of axionlike particles to nucleons
obtained from the Cavendish-type experiment (line labeled
"gr$_1$"), from measuring the minimum forces of
gravitational strength by means of torsion pendulum
(line labeled "gr$_2$"), from the Casimir-less experiment
(line 1), from measuring the Casimir pressure (line 2),
lateral Casimir force (line 3), and the Casimir force at
nanometer separations (line 4). The plane areas above
the lines are excluded and below them are allowed by the
measurement data.\label{fg3}}
\end{figure}
In Figure~\ref{fg3} we present in the logarithmic scale the strongest constraints
on the coupling constant $g$ as a function of the axion mass $m_a$. The line
labeled gr$_1$ is found \cite{36} from the Cavendish-type experiment \cite{70}
and the line labeled gr$_2$ \cite{25} from measuring the minimum forces of
gravitational strength by means of the planar torsional oscillators \cite{96,97,98}.
The line 1 is obtained \cite{94} from the Casimir-less experiment \cite{65},
line 2 from measuring the effective Casimir pressure \cite{60,61}, line 3 \cite{93}
from measuring the lateral Casimir force between corrugated surfaces \cite{52,53},
and line 4 \cite{51} from the experiment using a silicon carbide plate \cite{56,57}.
To summarize, the line gr$_1$ presents the strongest constraints on $g$ in the
region of axion masses $m_a<0.676~$meV, the line gr$_2$ in the region
$0.676~\mbox{meV}\leqslant m_a<4.9~$meV, the line 1 in the region
$4.9~\mbox{meV}\leqslant m_a<0.9~$eV, the line 2 in the region
$0.9~\mbox{eV}\leqslant m_a<8~$eV, the line 3 in the region
$8~\mbox{eV}\leqslant m_a<17.8~$eV, and the line 4 in the region
$m_a\geqslant 17.8~$eV.

{}From Figure~\ref{fg3}, it remains unclear for how long one could continue the
lines gr$_1$ and 4 to the left and to the right, respectively. The point is that in
the range of smaller and larger axion masses the strongest constraints on $g$
follow from some other types of laboratory experiments. They are sensitive to the
process of one-axion exchange between two nucleons and, thus, lead to the
spin-dependent effective potentials. Such kind of experiments are discussed in the
next section where we finally determine the current region of axion masses in
which the strongest constraints on $g$ follow from the Casimir and gravitational
experiments.

In the end of this section we briefly list some other spin-independent
potentials which are under discussion in the literature. As an example, the
spin-independent effective potentials between two massive spin-1/2 particles
arising due to an exchange of two spin-0 or two spin-1 massive bosons were found
in \cite{25} in the cases of scalar ($g_S$), pseudoscalar ($g_P$), vector ($g_V$),
and axial vector ($g_A$) couplings. The coupling (\ref{eq10}) considered above is
pseudoscalar, so that our coupling constant $g=g_P$ and the effective potential
(\ref{eq11}) is due to an exchange of two particles with pseudoscalar couplings.
In addition to (\ref{eq11}), the exchange of particles with one scalar and one pseudoscalar
couplings, as well as with one vector and one axial vector couplings, were considered
in \cite{25}. In so doing, in place (\ref{eq10}), the Lagrangian density with
the scalar coupling is given by
\begin{equation}
{\cal L}_{S}(x)=-g_S\bar{\psi}(x)\psi(x)
\varphi(x),
\label{eq14}
\end{equation}
\noindent
whereas the vector field of small mass $A_{\mu}$ is coupled to a fermion field
with an interaction of the following form
\begin{equation}
{\cal L}_{VA}(x)=\bar{\psi}(x)\gamma^{\mu}(g_V+g_A\gamma_5)\psi(x)
A_{\mu}(x),
\label{eq15}
\end{equation}
\noindent
where $\gamma^{\mu}$ with $\mu=0,\,1,\,2,\,3$ are the Dirac matrices.

Thus, the exchange of particles with one scalar and one pseudoscalar couplings between two
similar fermions results in the following effective potential \cite{25}
\begin{equation}
V_{SP}(r)=\frac{g_S^2g_P^2}{32\pi^2m r^2}\,e^{-2r/\lambda},
\label{eq16}
\end{equation}
\noindent
where $\lambda$ is the Compton wavelength of a spin-0 particle.
The effective potential due to an exchange of particles
with one vector and one axial vector
couplings is analogous to (\ref{eq16}) \cite{25}.

Some other spin-independent effective potentials have also been considered in
the literature. For example, such kind potentials arise between two fermions
due to the massive neutrino-antineutrino exchange where the direct coupling
to fermions is effected by the Z  bosons \cite{99}.
The possibilities of constraining the resulting intermolecular forces by means
of molecular spectroscopy are investigated in \cite{100}.

\section{Constraints on the Spin-Dependent Interactions}

As was already mentioned above, the originally introduced axions \cite{6,7} are
the pseudo Nambu-Goldstone bosons and their interaction with fermions is described
not by the Lagrangian density (\ref{eq10}) but by the pseudovector Lagrangian
density \cite{84,85}
\begin{equation}
{\cal L}_{PV}(x)=\frac{g}{2m_a}\bar{\psi}(x)\gamma_5\gamma_{\mu}\psi(x)
\partial^{\mu}\varphi(x).
\label{eq17}
\end{equation}
\noindent
The effective interaction constant in this Lagrangian density $g/(2m_a)$ is not
dimensionless and, as a consequence, the respective quantum field theory is
nonrenormalizable. On a tree level, however, both Lagrangian densities (\ref{eq10})
and (\ref{eq17}) result in the same action as can be seen performing the
integration by parts. Because of this, both (\ref{eq10}) and (\ref{eq17}) lead
to common spin-dependent effective potential caused by the exchange of one axion
between two fermions \cite{18,20,83,101}
\begin{equation}
V(r,\mbox{\boldmath$\sigma$}_1,\mbox{\boldmath$\sigma$}_2)=
\frac{g^2}{16\pi m^2}\left[(\mbox{\boldmath$\sigma$}_1\cdot\mbox{\boldmath$k$})
(\mbox{\boldmath$\sigma$}_2\cdot\mbox{\boldmath$k$})\left(
\frac{m_a^2}{r}+3\frac{m_a}{r^2}+\frac{3}{r^3}\right)-
(\mbox{\boldmath$\sigma$}_1\cdot\mbox{\boldmath$\sigma$}_2)\left(
\frac{m_a}{r^2}+\frac{1}{r^3}\right)\right]\,e^{-m_ar}.
\label{eq18}
\end{equation}
\noindent
Here, the same notations, as in Section 4, are used and
$\mbox{\boldmath$\sigma$}_1/2$, $\mbox{\boldmath$\sigma$}_2/2$ are the fermion spins
whereas {\boldmath$k$} is the unit vector
$\mbox{\boldmath$k$}=(\mbox{\boldmath$r$}_1-\mbox{\boldmath$r$}_2)/r$.

Unfortunately, all experiments of the Casimir physics performed up to date, as well
as gravitational experiments, deal with the unpolarized test bodies. As a result, the
interaction governed by (\ref{eq18}) averages to zero when integrated over their
volumes. For this reason, in Section~4 it was necessary to deal with the effective
potential (\ref{eq11}) which describes the two-axion exchange governed by the
Lagrangian density (\ref{eq10}). It is interesting that the effective potential of
an exchange by the two originally introduces axions, whose interaction with fermions
is described by the Lagrangian density (\ref{eq17}), is still unknown \cite{25}.
Because of this the results of Section~4 are only applicable to the GUT axionlike
particles. Below we consider two laboratory experiments which exploit the effective
potential (\ref{eq18}) and, thus, the constraints obtained from them pertain
equally to all types of axions and axionlike particles.

We begin with the comagnetometer experiment \cite{102} adapted for searches the
anomalous spin-dependent interaction between the mixture of K and ${}^3$He atoms
and the ${}^3$He spin  source spaced at a distance of 50~cm. In this experiment,
the energy resolution of $10^{-25}~$eV has been reached \cite{102} giving the
possibility to place strong constraints on the interaction potential (\ref{eq18})
with the coupling constant of axions to neutrons $g=g_P$.
Strong constraints were placed also on the effective potentials \cite{102}
\begin{equation}
V_1(r)=\frac{g_A^2}{4\pi r}\,
(\mbox{\boldmath$\sigma$}_1\cdot\mbox{\boldmath$\sigma$}_2)\,e^{-m_A r}
\label{eq19}
\end{equation}
\noindent
and
\begin{equation}
V_2(r)=-\frac{g_Vg_A}{4\pi m}
\big([\mbox{\boldmath$\sigma$}_1\times\mbox{\boldmath$\sigma$}_2]\cdot
\mbox{\boldmath$k$}\big)
\left(\frac{m_A}{r}+\frac{1}{r^2}\right)\,e^{-m_Ar},
\label{eq20}
\end{equation}
\noindent
which are caused by an exchange of light boson $A_{\mu}$ with the mass $m_A$
coupled to fermions by (\ref{eq15}).

\begin{figure}[!b]
\centering
\vspace*{-8.5cm}
\includegraphics[width=18 cm]{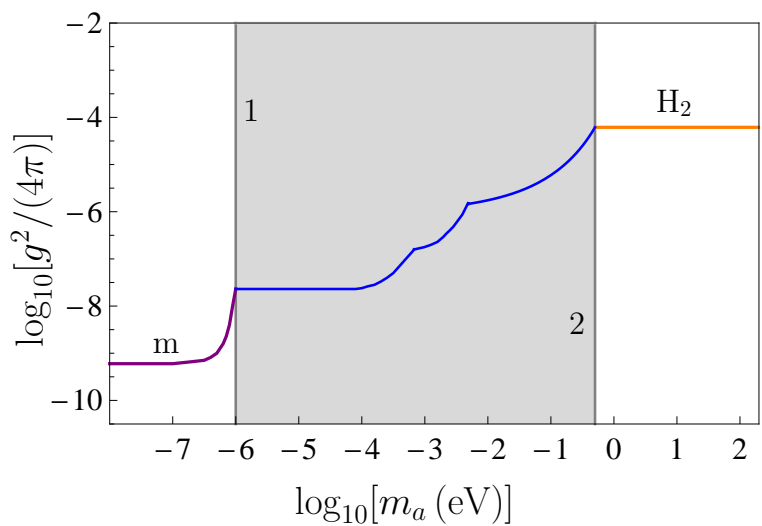}
\vspace*{-10.5cm}
\caption{The strongest laboratory constraints on the
coupling constant of axions and axionlike particles to
nucleons obtained from the magnetometer measurements
(line labeled "m") and from the experiment with a beam of
molecular hydrogen (line labeled "H$_2$"). The vertical
lines labeled 1 and 2 indicate the current borders
for the strongest constraints obtained from the
gravitational experiments and Casimir physics which
are shown by the envelope curve in between the lines 1
and 2. The plane areas above the lines are excluded
and below them are allowed by the measurement data.\label{fg4}}
\end{figure}
We illustrate the relative role of magnetometer measurements, on the one hand,
and the gravitational experiments and Casimir physics, on the other hand, in
constraining the axion-to-nucleon coupling in Figure~\ref{fg4} (recall that the
coupling constants of an axion to a proton and a neutron are assumed to be equal).
The line marked m in Figure~\ref{fg4} shows the constraints on the coupling
constant $g=g_P$ of the effective potential (\ref{eq18}) as a function of the
axion mass $m_a$ in the logarithmic scale obtained from the comagnetometer
measurements \cite{102}. The vertical line 1 at $m_a=1~\mu$eV indicates the
current border between the strongest laboratory constraints on the coupling
constant of axions to nucleons obtained from  the comagnetometer
measurements ($m_a<1~\mu$eV) and from the gravitational experiments and Casimir
physics ($m_a\geqslant 1~\mu$eV). At $m_a\geqslant 1~\mu$eV in Figure~\ref{fg4}
we plot the envelope curve of the strongest constraints on the coupling
constant of axions to nucleons from the gravitational experiments and Casimir
physics shown in Figure~\ref{fg3}.

In the region of larger axion masses the constraints on $g$ obtained from the
Casimir physics are greatly surpassed by the constraints found from the
experiment with a beam of molecular hydrogen \cite{103}. In this experiment,
the accurately measured dipole-dipole forces between two protons in the beam
of molecular hydrogen have been compared with theory. The constraints on the
coupling constant $g=g_P$ in the potential (\ref{eq18}) describing an exchange
of one axion between two nucleons were obtained from the measure of agreement
between experiment and theory. These constraints are shown by the line labeled
H$_2$ in Figure~\ref{fg4}. They remain the strongest ones with decreasing
axion mass down to $m_a=0.5~$eV. For smaller axion masses the strongest
constraints on $g$ follow from the Casimir physics and Cavendish-type
experiments. The vertical line~2 in Figure~\ref{fg4}
plotted at $m_a=0.5~$eV separates
the constraints on the coupling constant of axions to nucleons obtained from the
Casimir physics and from the interaction of protons in a beam of molecular
hydrogen. One can conclude that at the moment the gravitational experiments
and Casimir physics provide the strongest laboratory constraints on the
coupling of axionlike particles to nucleons over the wide range of axion masses
from $m_a=1~\mu$eV to 0.5~eV.

Note that the constraints of line~H$_2$ in Figure~\ref{fg4} remain the strongest
ones up to $m_a=200~$eV. For larger $m_a$ they are replaced with the stronger
constraints found from the comparison between the measurement data of the
magnetic resonance experiment and theory for the spin-spin interactions in
deuterated molecular hydrogen \cite{104}. Using the same approach, the strongest
constraints on the interaction constant $g_A$ of the effective potential (\ref{eq19})
were obtained in the region of masses of axial vector bosons
$m_A\leqslant 1000~$eV \cite{104}.

Very recently a Rb-${}^{21}$Ne comagnetometer using a SmCo$_5$ magnet as a spin
source was employed to obtain constraints on the products of the pseudoscalar
and axial vector electron and neutron couplings to light bosons $g_P^eg_P^n$
and  $g_A^eg_A^n$ \cite{105}. In so doing the coupling constants $g_P$ and $g_A$
belong to the spin-dependent effective potentials (\ref{eq18}) and (\ref{eq19}),
respectively, describing an interaction of the pseudoscalar and vector bosons
with either an electron or a neutron. As a result, the following constraints
have been obtained:
\begin{equation}
g_P^eg_P^n<1.7\times 10^{-14},
\qquad
g_A^eg_A^n<5\times 10^{-42},
\label{eq21}
\end{equation}
\noindent
which is valid for the boson masses below $1~\mu$eV \cite{105}.

Many other laboratory experiments on constraining the hypothetical interactions of
different types have been performed which are not directly relevant to the Casimir
physics. For instance, a constraint on the coupling constant of electron-electron
axial vector interaction $(g_A^e)^2/(4\pi)\leqslant 4.6\times 10^{-12}$
arising due to an exchange of spin-1 boson with masses $1~\mbox{eV}<m_A<20~$eV
was found in \cite{106} basing on measurements of the magnetic dipole-dipole
interaction between two Fe atoms \cite{107,108}. A stronger constraints
$(g_A^e)^2/(4\pi)\leqslant 1.2\times 10^{-17}$ within another range of boson masses
$m_A\leqslant 0.1~$eV was obtained \cite{109} from measuring the magnetic
interaction between two trapped ${}^{88}$Sr${}^{+}$ ions \cite{110}.
The coupling constant $g_A$ of the effective potential (\ref{eq19}) describing
the electron-muon interaction was constrained \cite{111} for the spin-1 boson
masses $m_A\sim 4~$keV by comparing the spectroscopic measurements with precise
QED calculations. Much work was also done on constraining the hypothetical
interactions which depend on both spins and velocities of the interacting
particles (see, e.g., \cite{112,113}).

The experimental tests considered in this section gave the possibility to determine
the region of parameters where the constraints obtained from the Casimir physics
are the strongest ones. It is possible, however, to constrain the spin-dependent
hypothetical interactions directly by measuring the Casimir interaction between
two polarized test bodies. In this case the additional interaction caused, e.g.,
by the effective potential (\ref{eq18}) is not averaged to zero after the
integration over their volumes and one can obtain the constraints on $g$ and $m_a$
from the measure of agreement between experiment and theory. The Casimir
experiments of this kind were proposed in \cite{114}. One option considered in
\cite{114} is to use the magnetic test bodies made, for instance, of the
ferromagnetic metal Ni (note that the Casimir interaction between two Ni-coated
test bodies was already measured in \cite{115,116} but Ni in that experiment was
not magnetized). It is significant that at the experimental distances
a magnetization of the test bodies results
in a spatially homogeneous magnetic force  which makes
no impact on the measured gradient of the Casimir force. This experiment could lead
\cite{114} to the constraints on the coupling constant of axions to electrons
which are by several orders of magnitude weaker than those already obtained for
solar axions produced by the Compton process and bremsstrahlung \cite{117}.

A more prospective possibility suggested in \cite{114} is to constrain the
interaction (\ref{eq18}) between nucleons by measuring the Casimir interaction
for the test bodies possessing the nuclear polarization. It has long been known that
spin polarization can be transferred from atomic electrons to nuclei.
For the silicon carbide test bodies (this material was already used in measurements
of the Casimir force discussed above \cite{56,57}) one can reach the 99\% polarization
of ${}^{29}$Si nuclear spins in silicon carbide by means of the  optical pumping
\cite{118}. Calculations show \cite{114} that measurements of the gradient of the
Casimir force between the silicon carbide test bodies possessing the nuclear
polarization could lead to stronger
constraints on the coupling constant of axions to nucleons for the axion masses
$m_a\sim 1~$eV.

\section{Constraints on Some Exotic Particles}

Light and massless hypothetical particles considered above and their interactions
with particles of the Standard Model are described by the conventional formalism
of local quantum field theory and respective effective potentials.
As mentioned in Section~1, one of the crucial problems of modern physics is the
problem of dark energy which contributes 68\% of the Universe energy and is
responsible for the acceleration of the Universe expansion \cite{4}.
In an effort to understand the structure of dark energy, several exotic particles
have been introduced in the literature whose properties are not unchanged, as for
all conventional elementary particles, but depend on the environmental conditions.

The first such particle is the so-called {\it chameleon} which is described by the
real self-interacting scalar field $\phi$ with a variable mass \cite{13}.
This particle becomes heavier, i.e., it has a shorter interaction range, in more
dense environments and lighter in the free space. The simplest field equation
for the static chameleon field can be written as \cite{13,119}
\begin{equation}
\Delta\phi=\frac{\partial V(\phi)}{\partial\phi}+\frac{\rho}{M}\,e^{\phi/M},
\label{eq22}
\end{equation}
\noindent
where $V(\phi)$ is the self-interaction potential, $M$ is the typical mass of the
background matter fields, and $\rho$ is the density of background matter.
An interaction with the background matter leads to a dependence of the chameleon
mass on $\phi$ \cite{119}
\begin{equation}
m(\phi)=m_0e^{\phi/M},
\label{eq22a}
\end{equation}
\noindent
where $m_0$ is the bare mass.

In the region of high matter density, for instance, on the Earth, $m(\phi)$ can be
rather large and, thus, the respective interaction sufficiently short-ranged to
avoid observable violations of the equivalence principle and the inverse square
law of Newtonian gravity. In the region of low matter density, however, e.g., in the
cosmic space, some new physics may appear.

There are several chameleon models depending on a specific form of the potential
$V(\phi)$ \cite{119}. For us it is important that an exchange of chameleons
between two test bodies leads to some additional force between them which can be
constrained in experiments on measuring the Casimir force. The character of this
force depends on both the chameleon model and on the experimental configuration.
In the configuration of two parallel plates and a sphere above a plate the
additional chameleon force was calculated in \cite{119}, and some constraints on
it following from the already performed experiments were found for different
types of the chameleon potential. In \cite{120} the new experiment was proposed
on measuring the Casimir force between two parallel plates at large separations
exceeding $10~\mu$m.  This allows to test different chameleon models by
varying the density of matter in a gap between the plates. The proposed
experiment was further elaborated and developed in \cite{80,81,123,124}.
Its realization is of much promise for this field.

The other kinds of exotic particles suggested in order to explain the nature of
dark energy are {\it symmetrons}. These particles are described by the
self-interacting real scalar field whose interaction constant with matter
depends on the density of an environment \cite{14,126,127}.
Symmetrons interact with matter weaker if the environment density is higher.
Similar to chameleons, this property helps them to escape notice in
gravitational experiments of E\"{o}tvos and Cavendish type.
In the static case the symmetron field satisfies the equation \cite{14,126}
\begin{equation}
\Delta\phi=\frac{\partial V(\phi)}{\partial\phi}+\
\left(\frac{\rho}{M^2}-\mu^2\right)\phi,
\label{eq23}
\end{equation}
\noindent
where $\mu$ is the symmetron mass and all other notations are the same as
in (\ref{eq22}).

Similar to the case of chameleons, there should be some additional force which
arises  between two closely spaced test bodies due to the exchange of
symmetrons. Because of this, measurements of the Casimir interaction can be
used for constraining the parameters of symmetron models. The forces arising
due to an exchange of symmetrons between two dense parallel plates separated
by a vacuum gap, as well as between a sphere and a plate, were found in
\cite{128}. The experiment using the sphere-plate geometry was proposed which
is capable to place strong constraints on the models of symmetron in near
future \cite{128}. In addition to the chameleons and symmetrons, the
properties of some other exotic particles, e.g., the dilaton introduced in
string theory, depend on the environment \cite{129}.

Explanation of the accelerated expansion of the Universe in the framework of
Einstein's general relativity theory requires the stress-energy tensor
describing the negative pressure of some material medium. Besides the
chameleon and symmetron fields, this property is offered by the stress-energy
tensor of the Maxwell-Proca electrodynamics in the case of nonzero photon
mass \cite{130}. An impact of the negative pressure originating from the
Maxwell stress-energy tensor of massive photons on the interstellar gas,
on stars in the process of their formation and on the rotational dynamics
of galaxies was investigated.

The effect of a nonzero photon mass on the Casimir-Polder interaction between
two polarized particles was considered in \cite{131} with applications to
several theoretical approaches beyond the Standard Model. It was noted that
the absence of deviations in the standard Casimir-Polder force calculated
using the massless photons from the measurement data could be used to place
new constraints on some extradimensional models.

The constraints on the so-called {\it hidden photons}, i.e., light spin-1 bosons,
which are predicted in the framework of string theory and do not interact
directly with elementary particles of the Standard Model, are obtained in
\cite{132} from experiments on measuring the Casimir force and testing the
Coulomb law. Further improvement of these constraints is expected in future
with increasing precision of relevant measurements.

\section{Implications for the Quantum Vacuum, Dark Energy, and the
Cosmological Constant}

One of the major unresolved problems of the Standard Model is the problem of
energy of the quantum vacuum. It is usually believed that the standard quantum
field theory should be applicable at all energies below the Planck energy
$E_{\rm Pl}=1/\sqrt{G}\sim 10^{19}~\mbox{GeV}\sim 10^9~$J.
The vacuum energy density of quantum fields is given by the divergent intergrals
with respect to momentum
\begin{equation}
\varepsilon_{\rm vac}=\frac{1}{2(2\pi)^2}\int d^3p\left(
\sum_{l=1}^{H}h_l\sqrt{m_l^2+{\mbox{\boldmath$p$}}^2}-
\sum_{l=1}^{F}f_l\sqrt{M_l^2+{\mbox{\boldmath$p$}}^2}\right).
\label{eq24}
\end{equation}
\noindent
Here, $H$ bosonic fields with masses $m_l$ and degrees of freedom $h_l$ and
$F$ fermionic fields with masses $M_l$ and degrees of freedom $f_l$ contribute
to the result. When making a cutoff at the Planck momentum
$p_{\rm Pl}=E_{\rm Pl}/c$, we obtain the huge energy density
$\varepsilon_{\rm vac}\sim 10^{111}~\mbox{J/m}^3$.
However, the vacuum energy density needed to explain the observed acceleration
of the Universe expansion is of the order of
$\varepsilon_{\rm vac}^{\rm obs}= 10^{-9}~\mbox{J/m}^3$ which is different from
the above value by the factor of $10^{120}$ \cite{133,134}.
Considering that the vacuum energy density can be interpreted in terms of the
 cosmological constant $\Lambda$ in Einstein's equations \cite{135}, so huge
 difference between its predicted and observed values is sometimes called
 {\it the vacuum catastrophe} \cite{2}.

In fact a consideration of the exotic particles and fields in Section~6, such as
chameleons, symmetrons, massive photons, etc., aims to explain the nature of a
background medium possessing the energy density of about $10^{-9}~\mbox{J/m}^3$
and contributing approximately 68\% to the total energy of the Universe.
To ensure the observed acceleration of the Universe expansion, this medium
should possess the negative pressure
$P_{\rm vac}^{\rm obs}=-\varepsilon_{\rm vac}^{\rm obs}<0$.
This property is guarantied by the cosmological term in Einstein's equations
\begin{equation}
R_{ik}-\frac{1}{2}Rg_{ik}+\Lambda g_{ik}=-8\pi G T_{ik},
\label{eq25}
\end{equation}
\noindent
where $R_{ik}$, $g_{ik}$, and $T_{ik}$ are the Ricci,  metrical, and
stress-energy tensors. The cosmological constant in (\ref{eq25}) is connected
with the observed vacuum energy density according to
\begin{equation}
\Lambda=8\pi G\varepsilon_{\rm vac}^{\rm obs}\approx
2\times 10^{-52}~\mbox{m}^{-2}.
\label{eq26}
\end{equation}

Taking into account that both the Casimir force and the cosmological constant
may have joint origin in the zero-point oscillations of the quantum vacuum, the
question arises about the possibility of their common theoretical description.
The point is that the Casimir effect is well understood in the framework of
the Lifshitz theory \cite{136,137}. In doing so the finite Casimir energy
density is obtained as a difference between two infinite quantities calculated
in the presence of a material medium with appropriate boundary conditions
and in the free space. The respective force was measured in numerous experiments
and found to be in agreement with theory \cite{44,45}.

Recently an attempt was undertaken to develop the Lifshitz-type theory of the
cosmological constant \cite{138}. This attempt is based on an observation
that the space-time geometry can be treated as an effective medium  for
the electromagnetic field. The medium under consideration
is characterized by some index of
refraction $n$ and the expansion of space is described as the time
dependence of $n=n(t)$. According to the argumentation of \cite{138},
the expanding flat space is indistinguishable from the uniform medium
having the refraction index depending on time. Under an assumption of similar
response to the electric and magnetic fields, $\varepsilon=\mu$ and, thus,
$n=\sqrt{\varepsilon\mu}=\varepsilon$, the contribution of the
electromagnetic field to the cosmological constant was found. This approach,
however, does not predict some specific value of $\Lambda$ because it
is considered not as a constant but as a dymanic quantity \cite{138}.
A comparison of the developed theoretical scheme with the experimental
data remains to be made.

Another approach considers the cosmological constant as one of the fundamental
constants of nature. Within this approach, it is not a problem that the bare
value of the cosmological constant $\Lambda_b=8\pi G\varepsilon_{\rm vac}$
is infinitely large. It is only important that its renormalized value is
given by (\ref{eq26}). From this point of view only the quantity
$\varepsilon_{\rm vac}^{\rm obs}$ represents the source of the gravitational
field whereas the infinitely large  $\varepsilon_{\rm vac}$, representing the
energy density of virtual particles, does not gravitate \cite{139,140}.
Although this approach could be finally approved only in the framework of
missing quantum gravity, it seems to be in close analogy with the Casimir effect
where only a difference between two infinite sets of the zero-point oscillations,
with and without  the material boundaries, contributes to the observable energy
density and force.

In any case, the tests of dark energy by means of atomic interferometry and
neutron scattering \cite{141,142} along with that ones considered above by means
of the Casimir and gravitational experiments should shed more light on this
puzzling form of matter.

\section{Discussion}

In the foregoing, we have considered different theoretical
predictions beyond the Standard Model and constraints on
them following from measurements of the Casimir force and
several other laboratory experiments which cover the parameter
regions neighboring to those covered by the Casimir effect.
It turns out that the predicted interactions and elementary
particles lead to a wide variety of effective potentials and
respective forces which could act between the test bodies in
addition to familiar fundamental interactions of the
Standard Model. During the last decades, a lot of experiments
have been performed attempting to discover these forces or at
least to constrain their parameters. Here, we restricted our
consideration to the constraints obtained from measuring the
Casimir force supplemented with the gravitational experiments,
neutron scattering, magnetometer measurements etc.

One of the most interesting subjects is the possibility of
new power-type interactions decreasing with separation
faster than the gravitational and electric ones. The
constraints on the potentials of the form $r^{-n}$ with
$n>1$ were obtained from the Cavendish-type experiments and
remained unchanged for more than a decade. The progress in
obtaining the stronger constraints reflected in Section 2
has been made only in 2020 for the potentials with $n=2,\,3,\,4$,
and 5 by using the measurement results of the further improved
Cavendish-type experiment. It seems that additional strengthening
of the obtained constraints may require new experimental
approaches.

The potentials and forces of Yukawa-type are of no less
importance. They are predicted in many theoretical schemes
beyond the Standard Model and were tested in numerical
experiments. Measurements of the Casimir force are the
recognized source of constraints on the Yukawa-type
potentials but there is a permanent competition between
different tests on whether or not this particular test leads
to the strongest constraints within some specific interaction
range. In Section 3, we have reflected the present situation
in this subject. Unlike the case of power-type potentials,
here the state of affairs changes rapidly and new important
results might be expected in near future.

The spin-independent potentials different from the Yukawa one
were considered in Section 4. The most important of them
describes an exchange of two axionlike particles between two
fermions. The constraints on the coupling constant and mass
of axionlike particles were obtained by using this process
from the Casimir effect and gravitational experiments. These
constraints compete between themselves and with the
constraints of another type which are obtained from the
experiments sensitive to an exchange of one axion. The more
exotic spin-independent potentials discussed by us may
become more usable in the coming years.

The exchange of one axion between two fermions is described
by the spin-dependent potential which is used for constraining
both the originally introduced axions and axionlike particles
by means of the magnetometer measurements, measurements of
dipole-dipole forces between two protons in the beams of
molecular hydrogen and in some other laboratory experiments.
These experiments are complementary to the Casimir physics
and gravitational measurements in constraining the
axion-nucleon interaction on a laboratory table. The
constraints on several more exotic spin-dependent potentials
under discussion in the literature were also considered in
Section 5. There are many suggestions of this kind and their
detailed consideration is beyond the scope of the present
review where we discussed only a few typical examples.

The hypothetical constituents of dark energy, such as
chameleons, symmetrons, massive photons etc., as well as the
constraints on them from different laboratory experiments
including measurements of the Casimir force, were discussed
in Section 6. The main problem here is somewhat indefinite
character of the theoretical predictions which admit
distinct forms of potentials and different values of the
involved parameters. The diverged approaches to a description
of the dark energy and the cosmological constant considered
in Section 7 demonstrate that we are still far from
understanding of their physical nature.

\section{Conclusions}

To conclude, the Casimir effect allows to place strong
constraints on many theoretical predictions beyond the
Standard Model including the spin-independent and
spin-dependent interactions and exotic particles and
fields introduced for an understanding of the nature of
dark energy. The borders of parameter regions, where
the Casimir effect leads to the strongest constraints,
are determined from some other laboratory experiments.
In near future one could expect that more important results
will be obtained using new generation of experiments in the
configuration of parallel plates spaced at separations
about $10~\mu$m and differential force measurements with
sensitivity below a femtonewton.
Because of this, it is very probable that testing and
constraining the far-reaching predictions of the
supersymmetry, supergravity, string theory and other
sophisticated theoretical approaches will be made not only
with the more powerful accelerators of next generations,
but with precision tabletop laboratory experiments.

\funding{This work was supported by the Peter the Great Saint Petersburg Polytechnic
University in the framework of the Russian state assignment for basic research
(project N FSEG-2020-0024).
}

\acknowledgments{The author is grateful to H. Abele, A.A. Banishev, V.B. Bezerra,
M. Bordag, R. Castillo-Garza, C.-C. Chang, H.C. Chiu, R.S. Decca, E. Fischbach,
D.E. Krause, P. Kuusk, D. L\'{o}pez, V.N. Marachevsky, U. Mohideen, V.M. Mostepanenko,
C. Romero, and R.I.P. Sedmik for collaboration in our joint publications used in
this review.}

\conflictsofinterest{The author declares no conflict of interest.}

\reftitle{References}

\end{document}